\documentstyle [12pt]{article}

\begin{document}

\begin{center}

{\large \bf A Rotating Quantum Vacuum} \\

\vspace{3cm}

V. A. De Lorenci\footnote{lorenci@lafex.cbpf.br} and 
N. F. Svaiter\footnote{nfuxsvai@lafex.cbpf.br}\\
\vspace{0.6cm}
{\it Centro Brasileiro de Pesquisas F\'{\i}sicas,} \\
\vspace{0.1cm}
{\it Rua Dr. Xavier Sigaud, 150, Urca} \\
\vspace{0.1cm}
{\it Rio de Janeiro CEP 22290-180-RJ. Brazil.} \\
\vspace{0.1cm}

\begin{abstract}

We investigate how a uniformly rotating frame is defined as the 
rest frame of an observer rotating with constant angular velocity 
$\Omega$ around the $z$ axis of an inertial frame. Assuming that 
this frame is a Lorentz one, we second quantize a free massless 
scalar field in this rotating frame and obtain that 
creation-anihilation operators of the field are 
not the same as those of an inertial frame. This leads to a new 
vacuum state --- a rotating vacuum --- which is a superposition of positive 
and negative frequency Minkowski particles. After this, introducing 
an apparatus device coupled linearly with the field we obtain that there 
is a strong correlation between number of rotating particles (in a given 
state) obtained via canonical quantization and via response function 
of the rotating detector. Finally, we analyse polarization effects in circular 
accelerators in 
the proper frame of the electron making a connection with the inertial
frame point of view. 

\end{abstract}

\end{center}

\vspace{1cm}

Pacs numbers: 04.62.+v, 03.65.Bz

\newpage

\section{Introduction}

\subsection{ Introductory Remarks}

The purpose of this paper is to discuss the puzzle of the 
rotating detector \cite{Dray} and to relate this to 
polarization effects of electrons in storage rings \cite{Bell}.  
We try to avoid many technical
difficulties to emphasize only fundamental results. 

The most important step in the development of general relativity
from special relativity is to accept the idea that is possible to
discuss physics --- and compare measurements --- not only between inertial
frames but also between any arbitrary frames of references.
It is natural to ask how to second quantize any field
in an arbitrary frame in Minkowski spacetime. Note that the
principle of general covariance require that physically observables
are always expressible in coordinate independent fashion. The development
of such ideas introduce a plethora of new phenomena. One of 
these is the Unruh-Davies effect \cite{effect,effect2}. 
An universal definition of 
the vacuum for a system described by a Hamiltonian is that 
the vacuum is the lowest energy state. If to describe the system 
we use a finite number of degrees of freedom, all 
representations of the operator's algebra are unitarily 
equivalents, i.e., different vacua lie in the same Hilbert space.
This means that the physical description of the system will 
not depend on the choice of representation.
However, if to describe the system we have to make use of infinite 
degrees of freedom, there are an infinite number of unitary 
inequivalent representations of the commutation 
relations \cite{Wightman}.
Different inequivalent representations will in general give rise to 
different pictures with different physical implications.

A well known example of this situation arises in the study of the 
quantization of a field by observers with linear proper 
acceleration \cite{Rindler}.
If we quantize a field in the Rindler's frame one finds quantization 
structure identical to the quantization obtained by inertial observers.
As in the case of inertial observers, in 
the Rindler's manifold there is a time-like Killing vector 
and the symmetry 
generated by this vector field is implemented by a unitary 
operator group.
The generator of this unitary group is positive definite and the 
construction of eingenstates of this operator allows a particle 
interpretation where a new vacuum state (the Fulling vacuum) is introduced \cite{Full}. The Minkowski and the Fulling vacua are non-unitarily 
equivalents. It is possible to 
show that the Minkowski vacuum can be expressed into a set 
of EPR type of Rindler's particles \cite{Lee}. As a natural consequence 
of this fact is that a particle detector at rest
in Rindler's spacetime interacting with a massless scalar field 
prepared in
the Minkowski vacuum responds as though is were at rest in 
Minkowski spacetime
immersed in a bath of thermal radiation.  Many authors claim 
that this case of linear acceleration is physically not very 
interesting 
since we need an eternal phase of constant acceleration.

A more treactable case (at least experimentally) is the case 
of transverse
acceleration found in circular movement. This particular situation 
introduce 
some interesting questions related with the meaning of particles in 
non-inertial frames of references. 
It has been sugested that the answer to the question: how would 
a particle detector responds in a given situation? can elucidate 
this problem. As we will see, this question introduce the rotating 
detector puzzle. To understand the problem of the rotating detector 
we have to go back 
to the problem of the rotating disc, i.e., the problem of rotation in 
relativity. A question that has interested many authors  
is whether the intrinsic geometry of a rotating 
disc is Euclidean or not. Infeld, using Einstein arguments \cite{Infeld} sustained 
that a rigid disc under uniform angular rotation $\Omega$ relative 
to an inertial frame will exhibit a non-Euclidean geometry (by 
a rigid body one understood a body in which during the motion no 
elastic stresses arises). The argument 
is that the circunference will suffer a Lorentz contraction 
although the 
radius $r$ will not. Consequently, the circunference of the 
rotating disc 
relative to an inertial frame is less than $2\pi r$. Lorentz had 
a opposed point of view \cite{Lorentz} and claimed that the intrinsic
geometry of the rotating disc is Euclidean since the radius 
and the circunference of the rotating disc contract by the 
same amount. 

A different approach to study this problem based on kinematic 
arguments has been presented by Hill long time ago \cite{Hill}. 
If the speed of any point in an uniform rotating 
disc is a {\em linear} function of the radius, distant points have
speeds exceeding the velocity of light. Hence this author  concluded that 
the speed-distance law must be non-linear and approach the velocity 
of light when the radius goes to infinity. Even today these are 
open questions and no definite answer has been given. 

The key point of the discusion  
is to find the correct transformation from an inertial
frame to a local frame in any point of the disk with constant angular
velocity $\Omega$. Let $\Sigma^{0} = (t^{'},\,r^{'},\,\theta^{'},\,z^{'})$ 
be an
inertial frame with cylindrical coordinates $r^{'},\,\theta^{'},\,z^{'}$ and
the common frame time $t^{'}$ measured by synchronised standard clocks.
Consider a point $P$ revolving around the $z^{'}$ axis at fixed distance
$r^{'}_P$ with constant angular velocity $\Omega_P$ with respect to 
$\Sigma^{0}$, i.e.,
\begin{eqnarray}
\frac{{\rm d}\theta^{'}_P}{{\rm d}t^{'}} &=& \Omega_P
\label{eqP} \\
\frac{{\rm d} r^{'}_P}{{\rm d}t^{'}} &=& \frac{{\rm d}z^{'}_P}{{\rm d}t'} = 0.
\label{rz}
\end{eqnarray}
In the rest frame of $P$, $\Sigma_P$ we introduce the coordinates
$t^{(r')},\,r,\,\theta^{(r')}$ and $z$ with $z^{'}=z$ and $r^{'}=r$.
Note that $t^{(r')}$ is the local time variable at any point $r=r^{'}$ and
$\theta^{(r')}$ is the local angular variable at any point $r=r^{'}$. 
If someone assume that the rotating frame 
is a Galilean one and quantize a massless scalar field in such frame,
it is possible to show that the rotating vacuum is just the Minkowski vacuum.
The mistake lies in the fact of use a Galilean tranformation for
the introduction of coordinates in the rotating frame.
The price we have to pay is that 
an aparatus device (a detector coupled with a field) which gives 
information about the particle content of the state of the field 
can be excited if it is prepared in the ground state with the 
field in the Minkowski vacuum \cite{Letaw}. This is an odd result.
One would expect the rotating detector {\it not to be excited} 
by the rotating vacuum. In this paper we will try to sheed some light 
on these problems.

We would like to stress that we are not interested here in discussing the 
subtle problem of how to decode the information stored in the 
composite system (detector and scalar field) to convert in a classical 
sign. The modern treatment of this problem is the following: both 
the detector and the scalar field are not closed systems but they are 
open systems interacting with the enviroment. In this way certain 
phase relations disappear, i.e., loss of coherence to its enviroment 
(Decoherence). This idea allows that the composite system (detector 
and the scalar field) be described by a diagonal 
matrix density \cite{Zurek}.
For an application of such ideas in the Unruh-Davies 
effect see for example 
Ref. \cite{Muller}. 

\subsection{Synopsis}

The paper is organized as follows. In section 2 we discuss how a
uniformly rotating frame is defined as the rest frame of an observer
rotating with constant angular velocity $\Omega$ around the $z'$ axes of an
inertial frame $\Sigma^{0}$. We first assume that the rotating frame 
is a Galilean one and second quantize a massless scalar field in such frame.
We show that the rotating vacuum is just the Minkowski vacuum.
Some disturbing situations are analysed.
In section 3 we discuss radiative 
processes assuming that a uniformly rotating frame must be a Lorentz
frame, i.e., we have to use a Lorentz-like transformation for the 
introduction of coordinates in the rotating frame. 
It is shown that the resulting rotating vacuum of the canonical quantization 
framework is not the Minkowski vacuum. 
In section 4 we perform the second quantization of the total
Hamiltonian of the system to show that the process of 
an absorption (emission) 
of a rotating particle and excitation (decay) of the detector
in the non inertial frame is interpreted as 
an excitation (decay) of the detector  with emission of 
a Minkowski particle
in the inertial frame. Conclusions are given in section 5.
In this paper we use $\hbar=c=1$.

\section{ The Rotating Frame Candidates}

The problem of the rotating disc have been investigated 
by many authors and can be posed in the following way. 
Suppose the Minkowski spacetime with line element in the 
cylindrical coordinate system $x^{'\mu}=(t',r',\theta',z')$ adapted 
to an inertial frame given by
\begin{equation}
{\rm d}s^{2}={\rm d}t'^{2}-{\rm d}r'^{2}-r'^{2}{\rm d}\theta'^{2}-{\rm d}z'^{2}.\label{1}
\end {equation}
Suppose a disc rotating uniformly about the z axis with angular velocity
$\Omega$. How a uniformly rotating frame is defined, i.e., the rest frame
of an observer rotating with constant angular velocity $\Omega$ around the
$z'$ axes of an inertial frame $\Sigma^{0}$? In order words, we have to find
the correct transformations formula for the transition $\Sigma^{0} \rightarrow
\Sigma^{r}$, where $\Sigma^{r}$ is the local frame at any point of the disk
with angular velocity $\Omega$.
Which mapping we have to assume to compare measurements made 
in an inertial and in a rotating frame of reference? 
Eddington \cite{Eddington}, Rosen \cite{Rosen}
and Landau and Lifshitz \cite{Landau} adopted a coordinate system 
adapted to the rotating disk in such that transformation law between
the cylindrical coordinate system  $x'^\mu =(t',r',\theta',z')$ adapted to an
inertial frame and rotating coordinate system $x^\mu =(t,r,\theta,z)$
adapted to the disk are given by:
\begin{eqnarray}
t&=&t', \label{2}
\\
r&=&r', \label{3}
\\
\theta&=&\theta'-\Omega t', \label{4}
\\
z&=&z'. \label{5}
\end{eqnarray}
In the rotating coordinate system $x^{\mu}=(t,r,\theta,z)$ the 
line element
can be written as
\begin{equation}
ds^{2}=(1-\Omega^{2}r^{2})dt^{2}-dr^{2}-r^{2}d\theta^{2}-dz^{2}
+2\Omega r^{2}d\theta dt.\label{6}
\end{equation}
The line element in 
the rotating frame is stationary but not static. The world line 
of a point 
of the disc is an integral curve of the Killing vector 
$\xi=(1-\Omega^{2}r^{2})^{-1/2}\partial/\partial t$
which is timelike only for $\Omega r<1$. Rosen claimed that using the 
transformations given by eqs.(\ref{2}-\ref{5}) the 
speed-distance law is linear and this put a limit on the 
size of the disc that rotate with a given angular velocity. The 
same point of view was given by Landau and Lifshitz.

A ``more natural'' way to investigate such problem is to looking for a 
Lorentz frame such that the velocity of the disk does not obey a linear 
velocity law. In this case we have to find a 
naturally adapted coordinate system to this infinite rotating disk without 
the problems found in the Landau's {\it et al} coordinate system.

A second possibility trying to avoid the disc problem in the core of the 
discussion is to follow Hill's arguments.
This author presented a different answer for the problem. He raised 
the question if it is possible to find a group of transformation 
between the inertial and the non-inertial frame in such a way that for
small velocities we obtain the linear speed-distance law and for large 
distance approach the speed of light. Such a
transformation frames was presented by 
Trocheries \cite{Troch} and also Takeno \cite{Takeno}. In Takeno\rq s
derivation three assumptions are used: 
the transformation law constitute a group, for small velocities we 
recovered the usual linear velocity law ($v = \Omega r$) and velocity
composition law is also in agreement with special relativity. One 
hopes that the transformation law derived assuming this rules is
unique, although the proof of such assumption is missing.  

The coordinate transformations derived by Takeno are given by
\begin{eqnarray}
t&=&t'\cosh\Omega r' -r'\theta'\sinh\Omega r',\label{7}
\\
r&=&r',\label{8}
\\
\theta&=&\theta'\cosh\Omega r' - \frac{t'}{r'}\sinh\Omega r',\label{9}
\\
z&=&z'.\label{10}
\end{eqnarray}
Note that if we assume this mapping to connect measurements 
made in the rotating frame and those made in the inertial
frame, in the rotating coordinate system the line element 
assume a non-stationary form
\begin{equation}
{\rm d}s^{2}={\rm d}t^{2}-(1+P){\rm d}r^{2}-r^{2}{\rm d}\theta^{2}-{\rm d}z^{2}
+2Q{\rm d}r{\rm d}\theta+2S{\rm d}r{\rm d}t,\label{11}
\end {equation}
where $P$, $Q$ and $S$ are given by
\begin{eqnarray}
P&=&(\frac{Y}{r^{2}}+4\Omega\theta t)\sinh^{2}\Omega r -
\frac{\Omega}{r}(t^{2}+r^{2}\theta)\sinh^{2}2\Omega r +
\Omega^{2}Y,\label{12}
\\
Q&=&r\theta\sinh^{2}\Omega r-\frac{1}{2}t\sinh 2\Omega r+
\Omega rt,\label{13}
\\
S&=&\frac{t}{r}\sinh^{2}\Omega r-\frac{1}{2}\theta\sinh 2\Omega r-
\Omega r \theta,\label{14}
\end{eqnarray}
and
\begin{equation}
Y=(t^{2}-r^{2}\theta^{2}).\label{15}
\end{equation}
We would like to remark that this above line element define the intrinsic 
geometry of the rotating disk. Furthermore this line element define a 
multiple connect manifold. We will discuss this point further.

Before starting to analise the detector problem we would like to 
present some experimental and theoretical arguments against and in favour
of Trocheries and Takeno's coordinate transformation. The 
Special Theory of Relativity
show us that different inertial frames are connected by 
Lorentz transformations. Why we use a Lorentz-like transformation 
to connect measurements in the inertial and the non-inertal frame? 
We should mention that it is possible to write the transformations 
defined by eqs.(\ref{7}-\ref{10}) 
making a analogy with the Lorentz transformations.
Let us define $l=r\theta$ and $\gamma=(1-v^{2})^{-1/2}$. 
It is straightforward to show that eq.(\ref{7}) and eq.(\ref{9}) becomes
\begin{equation}
t=\gamma(t'-vl')\label{16}
\end{equation}
and
\begin{equation}
l=\gamma(l'-vt').\label{17}
\end{equation}
In other words the transformations defined by Trocheries and Takeno are
``Lorentz-like" transformation. The fundamental 
difference is that in this case the velocity is $v=\tanh\Omega r'$. 
It has been sugested by Phipps \cite{Phipps} that the Takeno's velocity 
distribution does not agree with the experimental data. 
Strauss \cite{Strauss} also adopted a Lorentz-like transformation,
but with a linear $v=\Omega r$ speed-distance law. The important 
consequence is that the light velocity on the rotating frame is one. 
Again, some authors claim that this result is in contradiction with 
the Sagnac's effect \cite{Sagnac,Gron}. The only way
to have results consistents with 
this effect is to use a ``Galilean" transformation given by eqs.(\ref{2}-\ref{5}). 
We would like to stress that the above arguments does not establish 
conclusively that we have to use the Galilean transformations.
As we will show, direct supports of Lorentz-like transformation
between both frames are suplied by the rotating detector puzzle and the
depolarization effect of electrons in a circular accelerator.

To investigate the meaning of particle in an arbitrary frame 
in a flat spacetime we have two different routs. The first is to canonical 
quantize the field and obtain the number of particles
operator for each mode $N_{R}(\omega)=b^{\dag}(\omega)b(\omega)$ in the 
arbitrary frame. 
For static line elements (Rindler, for example) this can be 
done in a unambiguosly way. For time dependent line elements (Milne, for example) 
it is possible to define instantaneous positive and negative frequency 
modes and diagonalize an instantaneous Hamiltonian operator.
The second rout is to introduce an measuring 
device, i.e., a detector (atoms) 
with  a coupling with the field. Experimentalists detect photons in 
laboratories. They are absorbed at fixed instants and 
cause the electrons 
in the atoms to jump from a ground state to an excited state. 
Glauber an others produced a theory of photo-detection 
using the rotating-wave approximation (RWA). In this 
approximation the detector (square-law detector) must gives information 
about the particle content of some state \cite{Glauber,Allen}. 
In other words, square-law 
detectors goes to excited state by absorption of quanta
of the field. 

Before continue let us discuss some arguments pro and contrary 
of the Glauber's detector. Bykov and Takarskii \cite{Bykov} showed 
that this detector model violates the causality principle 
for short observations times. If we assume that the observation time is 
large compared with $E^{-1}$, everething is in order. 
Note that 
it is possible to consider measurements of finite 
durations only for 
$\Delta T>1/E$. Of short time intervals we cannot even 
define the 
two-level system. Nevertheless there are some situations were we 
can not use the RWA, for example in resonant interaction between two 
atoms \cite{atoms}. As we will see the RWA can not be used only 
to find the rate of spontaneous decay. The same situation occur in a 
semi-classical theory of spontaneous emission where an atom in the 
excited state is stable since there is no vaccum fluctuations.

Going back to our problem, let us discuss these two routs 
that are usually used to investigate the meaning of particles 
in a arbitrary frame of reference. 
Let first perform the quantization of a massless real scalar 
field in the Landau's rotating frame. 
First we have to solve the Klein-Gordon equation 
in the $x^{\mu}=(t,r,\theta,z)$ coordinate system given by
\begin{equation}
\left[(\frac{\partial}{\partial t}-
\Omega\frac{\partial}{\partial\theta})^{2}-
\frac{1}{r}\frac{\partial}{\partial r}r\frac{\partial}{\partial r}
-\frac{1}{r^{2}}\frac{\partial^{2}}{\partial\theta^{2}}-
\frac{\partial^{2}}{\partial z^{2}}\right]\varphi(t,r,\theta,z)=0 \label{18}
\end{equation}
to find the normal modes that satisfies 
\begin{equation}
L_{\bar{K}}u_{\bar{q}mk_{z}}(t,r,\theta,z)=-i\bar{\omega} u_{\bar{q}mk_{z}}(t,r,\theta,z), \label{19}
\end{equation}
where $\bar{K}$ is a time-like Killing vector.
It is not difficult to show that the modes are 
given by \cite{Letaw,Denardo}
\begin{equation}
u_{\bar{q}mk_{z}}(t,r,\theta,z)=\frac{1}{2\pi\left[2(\bar{\omega}+m\Omega )\right]^{\frac{1}{2}}}
e^{-i\bar{\omega} t}e^{im\theta}e^{ik_{z}z}J_{m}(qr) \label{20}
\end{equation}
where
\begin{equation}
(\bar{\omega}+m\Omega)^{2}=k_{z}^{2}+q^{2},\label{21}
\end{equation}
\begin{equation}
(\bar{\omega}+m\Omega)>0,\label{22}
\end{equation}
and the radial part $J_{m}(qr)$ are the Bessel functions of first kind \cite{Lebedev}.
To continue the canonical quantization, the field operator $\varphi(t,r,\theta,z)$ have to be expanded using these
modes and the complex conjugates $\{u_{\bar{q}mk_{z}}(t,r,\theta,z),
u_{\bar{q}mk_{z}}^{*}(t,r,\theta,z)\}$, i.e.,
\begin{equation}
\varphi(t,r,\theta,z)=
\sum_{m}\int \bar{q}{\rm d}\bar{q}{\rm d}k_{z}\left[a_{\bar{q}mk_{z}}u_{\bar{q}mk_{z}}(t,r,\theta,z)+
a_{\bar{q}mk_{z}}^{\dag}u_{\bar{q}mk_{z}}^{*}(t,r,\theta,z)\right] \label{23}.
\end{equation}
Of course, in stationary geometries the definition 
of positive and negative frequency modes 
has no ambiguities. To compare both quantizations
i.e., in the inertial and in the rotating frame, we have to find the 
Bogoliubov coefficients between the inertial modes (cilindrical waves)
$\{\psi_{k}(t',r',\theta', z'),\psi_{k}^{*}(t',r',\theta', z')\}$
and the non-inertial ones given by eq.(\ref{20}). Since the Bogoliubov
coefficients $\beta_{k\nu} = - (u_{qmk_{z}},\psi_{k})$ 
are zero, Letaw and Pfautsch concluded that 
the vacuum defined by
\begin{equation}
a(\bar{q},m,k_{z})|0,R>=0 \qquad\forall\qquad \bar{q},m,k_{z},\label{24}
\end{equation}
i.e., the rotating vacuum is just the Minkowski vacuum. 
Note that we are not interested to discuss complications 
introduced by infinite-volume divergences. To circunvented this 
problem the  
creation and anihilation operators have to be smeared with 
square integrable test functions (wave-packet).

The introduction of the detector in this quantization scheme 
raised a fundamental question. If we prepare a detector in the ground state 
and the field in the Minkowski vacuum there is a non-null 
probability to find the detector in the excited state if the 
detector travel in a rotating world line, parametrized 
by eqs.(\ref{2}-\ref{5}). 
The orbiting detector will ``measure" quanta of the field although 
there is no rotating quanta in the Minkowski vacuum.  
How is 
possible to a detector being excited if is traveling in a
rotating disc if we prepare the field in the Minkowski vacuum? 
After the absorption, the field will be in a lower energy 
level than the
``original vacuum''. Therefore this ``original vacuum'' is 
not the true vacuum of the field.
Another way to formulate the problem is the following one:
our physical intuition say that a a rotating particle detector 
in the ground state interacting with the scalar field prepared in the
rotating vacuum must stay in the ground state.
Nevertheless, assuming the Galilean frame, the 
Minkowski vacuum $|0,M>$ is exactly 
the rotating vacuum $|0,R>$ and the rate of excitations 
instead to be zero 
is different from zero. The detector behaves as if it is not coupled to the 
vacuum, concluded Davies, Dray and Manogue \cite{Dray}. 
This is the so called
rotating detector puzzle.    
Some time ago 
Grove and Ottewill trying to sheed some light for these problem studied 
extended detectors \cite{Grove}.  
Letaw and Pfautsch, Padmanabhan \cite{Padmanabhan} and also 
Padmanabhan and Singh \cite{Singh}
concluded that the correlation between 
vacuum states defined via canonical quantization and via 
detector is broken in this particular situation. 
We cannot agree with this conclusion. 
The preceding considerations suggest that we can not accept a 
Galilean rotating
frame with a maximum radius given by 
($R_{\mbox{{\footnotesize max}}}=1/\Omega$).
In the next section we will remember the formalism and discuss
some possibilities to solve the rotating detector puzzle and the 
interpretational difficulties associated with it.

\section{Radiative Processes of the Monopole 
Detector and a New Rotating 
Vacuum}

Let us consider a system (a detector) endowed with internal degrees 
of freedom defining two energy levels with energy $\omega_{g}$ and 
$\omega_{e}$, $(\omega_{g}< \omega_{e})$ and respective 
eigenstates $\left|g\right>$ and $\left|e\right>$ 
\cite{effect2,Unruh,DeWitt}. 
This system is weakly coupled with a 
hermitian massless scalar field $\varphi(x)$ with interaction 
Lagrangian 
\begin{equation}
L_{int}=\lambda m(\tau)\varphi(x(\tau)),\label{25}
\end{equation}
where $x^{\mu}(\tau)$ is the world line of the detector 
parametrized using the proper time $\tau$, $m(\tau)$ is the 
monopole operator of the detector and $\lambda$ is a small 
coupling constant between the detector and the scalar field. 
For different 
couplings between the detector and the scalar field 
see for example 
Ref.\cite{Hinton} and also Ford and Roman \cite{Ford2}.

In order to discuss radiative processes of the whole system
(detector plus the scalar field), let us define the Hilbert
space of the system as the direct product of the Hilbert space
of the field $\bf H_{F}$ and the Hilbert space of the detector
$\bf H_{D}$
\begin{equation}
\bf H=\bf H_{D}\otimes \bf H_{F}. \label{26}
\end{equation}
The Hamiltonian of the system can be written as:
\begin{equation}
H=H_{D}+H_{F}+H_{int} \label{27},
\end{equation}
where the unperturbed Hamiltonian of the system is composed of 
the noninteracting detector Hamiltonian $H_{D}$ and the free 
massless scalar field Hamiltonian $H_{F}$.
We shall define the initial state of the system as:
\begin{equation}
\left|{\cal T}_{i}\right>=\left|j\right>\otimes\left|\Phi_{i}\right>, \label{28}
\end{equation}
where $\left|j\right>$, $(j=1,2)$ are the two 
possible states of the detector
($\left|1\right>=\left|g\right>$ 
and $\left|2\right>=\left|e\right>)$ 
and $\left|\Phi_{i}\right>$ is the initial state 
of the field. In the interaction picture, the evolution of the
combined system is governed by the Schrodinger equation
\begin{equation}
i\frac{\partial}{\partial\tau}\left|{\cal T}
\right>=H_{int}\left|{\cal T}\right>, \label{29}
\end{equation}
where
\begin{equation}
\left|{\cal T}\right>=U(\tau,\tau_{i})\left|{\cal T}_{i}\right>, \label{30}
\end{equation}
and the evolution operator $U(\tau,\tau_{i})$ obeys
\begin{equation}
U(\tau_{f},\tau_{i})=1-i\int_{\tau_{i}}^{\tau_{f}}H_{int}
(\tau^{'})U(\tau^{'},\tau_{i})d\tau^{'}. \label{31}
\end{equation}
In the weak coupling regime, the evolution operator can be 
expanded in power series of the interaction Hamiltonian.
To first order, it is given by
\begin{equation}
U(\tau_{f},\tau_{i})=1-i\int_{\tau_{i}}^{\tau_{f}}{\rm d}\tau^{'}
H_{int}(\tau^{'}). \label{32}
\end{equation}
The probability amplitude of the transition from the initial 
state $\left|{\cal T}_{i}\right>
=\left|j\right>\otimes\left|\Phi_{i}\right>$ at the 
hypersurface $\tau=0$ to 
$\left|j^{'}\right>\otimes\left|\Phi_{i}\right>$ at $\tau$ is given by
\begin{equation}
\left<j^{'}\Phi_{f}\right|U(\tau,0)\left|j\Phi_{i}\right>
=-i\lambda\int_{0}^{\tau}{\rm d}\tau^{'}\left<j^{'}\Phi_{f}\right|m(\tau^{'})
\varphi(x(\tau^{'}))\left|j\Phi_{i}\right>, \label{33}
\end{equation}
where $\left|\Phi_{f}\right>$ is an arbitrary state of the field and 
$\left|j^{'}\right>$ is the final state of the detector.
The probability of the detector being excited at the 
hypersurface $\tau$, assuming that the detector was prepared 
in the ground state is:
\begin{equation}
P_{eg}(\tau)=\lambda^{2}|\left<e\right|m(0)\left|g\right>|^{2}
\int_{0}^{\tau}{\rm d}\tau^{'}\int_{0}^{\tau}{\rm d}\tau^{''}
e^{iE(\tau^{''}-\tau^{'})}\left<\Phi_{i}\right|\varphi(x(\tau^{'}))
\varphi(x(\tau^{''}))\left|\Phi_{i}\right>, \label{34}
\end{equation}
where $E = \omega_{e}-\omega_{g}$ is the energy gap 
between the eigenstates 
of the detector.
Note that we are interested in the final state of the detector
and not that of the field, so we sum over all the possible
final states of the field $\left|\Phi_{f}\right>$. Since the 
states are
complete, we have
\begin{equation}
\sum_{f}\left|\Phi_{f}\right>\left<\Phi_{f}\right|=1. \label{35}
\end{equation}
Eq.(\ref{34}) shows us that the probability of excitation 
is determined by an integral transform of the positive Wightman function.

Before starting to analyze radiative processes, we would 
like to point out
that a more realistic model of particle detector must also have 
a continuum of 
states. This assumption allows us to use a first 
order perturbation
theory without taking into account higher order corrections. 
Although
we will use in this paper the two-state model, the case of a 
mixing between a discrete and a continuum 
eigenstates deserves further investigations.
For a complete discussion of the detector problem see for example
Refs. \cite{Ginzburg,Takagi,Sciama}.
In this section we will use a different notation. Two distincts spacetime 
points in the rotating coordinate system will be given by 
$x^{\mu}=(\eta,\xi)$ and  
$x'^{\mu}=(\eta',\xi')$. We are using the variable $\xi$ to represent both 
coordinates $r$ and $\theta$ i.e. $\xi\equiv \{r,\theta\}$. In the applications 
to storage ring we will be not interested in the $r$ and $\theta$ dependence of
the probability transition and will use only $P_{12}(E,\Delta T)$.
Since we are interested in 
finite time measurements let us follow Svaiter and Svaiter \cite{Nami1},  
and also Ford, Svaiter and Lyra \cite{Lyra} defining 
\begin{equation}
\eta-\eta{'}=\zeta \label{36}
\end{equation}
and
\begin{equation}
\eta_{f}-\eta_{i}=\Delta T.\label{37}
\end{equation}
We would like to stress that Levin, Peleg and Peres 
\cite{Peres} also used the same 
technique to study radiative processes in finite observation times.
Substituting eq.(\ref{36}) and eq.(\ref{37}) in eq.(\ref{34}) and defining 
$F(E,\Delta T,\xi,\xi')$ by 
\begin{equation}
P_{12}(E,\Delta T,\xi,\xi')= 
\lambda^{2}|\left<2\right|m\left|1\right>|^{2}
F(E,\Delta T,\xi,\xi') \label{38}
\end{equation}
we have
\begin{equation}
F(E,\Delta T, \xi,\xi')=
\int^{\Delta T}_{-\Delta T}{\rm d}\zeta(\Delta T-|\tau|)
e^{iE\zeta}
\left<0,M\right|\varphi(\eta^{'},\xi^{'})
\varphi(\eta,\xi)\left|0,M\right>. \label{39}
\end{equation}
It is clear that $F(E,\Delta T,\xi,\xi')$
is the probability of 
excitation normalized by the selectivity of the detector. 
The same can be done for decay processes and the probability of decay 
$P_{21}(E,\Delta T,\xi,\xi')$ is given by 
\begin{equation}
P_{21}(E,\Delta T,\xi,\xi')= \lambda^{2}|\left<1\right|m\left|2\right>|^{2}
F(E,\Delta T,\xi,\xi').\label{40}
\end{equation}
Let us define the rate $R(E,\Delta T,\xi,\xi')$, i.e., 
this probability transition per unit proper time as:
\begin{equation}
R(E, \Delta T,\xi,\xi')=\frac{d}{d(\Delta T)}F(E,\Delta T,\xi,\xi'). \label{41}
\end{equation}
Writting in a concise form we have:
\begin{equation}
R(E,\Delta T,\xi,\xi')=
\int^{\Delta T}_{-\Delta T}{\rm d}\zeta
e^{iE\zeta}
\left<0,M\right|\varphi(\eta^{'},\xi^{'})\varphi
(\eta,\xi)\left|0,M\right>. \label{42}
\end{equation}

This important result shows that asymptotically the rate of 
excitation (decay) of the detector is given by the Fourier transform 
of the positive frequency 
Wightman function. This is exactly the quantum version of the 
Wiener-Khintchine theorem which asserts that the spectral density 
of a stationary random variable is the Fourier transform of the 
two point-correlation function. 
Spliting the field operator in positive and negative 
frequency parts, the
rate becomes:
\begin{eqnarray}
R(E,\Delta T,\xi,\xi')&=&
\int^{\Delta T}_{-\Delta T}{\rm d}\zeta
e^{iE\zeta}\left[
\left<0,M\right|\varphi^{(+)}(\eta^{'},\xi^{'})
\varphi^{(+)}(\eta,\xi)\left|0,M\right>\right.
\nonumber\\ 
&&+
\left<0,M\right|\varphi^{(-)}(\eta^{'},\xi^{'})
\varphi^{(-)}(\eta,\xi)\left|0,M\right>
\nonumber\\ 
&&+
\left<0,M\right|\varphi^{(-)}(\eta^{'},\xi^{'})
\varphi^{(+)}(\eta,\xi)\left|0,M\right>
\nonumber\\ 
&&+
\left.\left<0,M\right|\varphi^{(+)}(\eta^{'},\xi^{'})
\varphi^{(-)}(\eta,\xi)\left|0,M\right>\right] . \label{43}
\end{eqnarray}
The last matrix element can be writen as                 
\begin{eqnarray}
\left<0,M\right|\varphi^{(+)}(\eta',\xi')
\varphi^{(-)}(\eta,\xi)\left|0,M\right>&=&
\left<0,M\right|\varphi^{(-)}(\eta,\xi)
\varphi^{(+)}(\eta',\xi')\left|0,M\right>\nonumber\\ &+&
[\varphi^{(+)}(\eta',\xi'),\varphi^{(-)}(\eta,\xi)]. \label{44}
\end{eqnarray}
The commutator is a c-number independent of the initial 
state of the field. Many authors in quantum optics 
claim that this contribution has no great physical 
interest. So the matrix elements determining the 
detection of quanta of the field are of the form 
\begin{equation}
\left<0,M\right|\varphi^{(-)}(\eta^{'},\xi^{'})
\varphi^{(+)}(\eta,\xi)\left|0,M\right>
 + \left<0,M\right|\varphi^{(-)}(\eta,\xi)
\varphi^{(+)}(\eta',\xi')\left|0,M\right>. \label{45}
\end{equation}
Substituting the modes given by eq.(\ref{20}) in eq.(\ref{42}) it
is possible to show that the rotating detector has non-zero 
probability of excitation. Since the contribution given by eq.(\ref{45})
is zero (there are no rotating particles in Minkowski vacuum), the 
non-zero rate is caused by the last term in eq.(\ref{44}). 
A disagreable situation emerges.
Our apparatus device is not measuring the particle 
content of some state.

The first solution of the puzzle of the rotating detector
was given a few months ago by Davies, Dray and Manogue \cite{Dray}. 
These authors assumed that the field is defined only in the 
interior of a cilinder of radius $R$ in such a way that the 
rotating Killing vector $\partial_{t}-\Omega\partial_{\theta}$ is 
always timelike. Consequently the response function of the rotating 
detector is zero. Of course if the angular velocity of the 
detector is above some threshold, excitation occurs. Clearly the 
excitation of the rotating detector is related with the mistake of 
use a Galilean transformation for the introduction of coordinates
in the rotating frame.

Let us assume that a uniformly rotating frame must be a Lorentz frame.
In this situation a naturally adapted coordinate system to a such frame
is the one defined by Trocheries and Takeno \cite{Troch,Takeno}. 
The advantage of this coordinate system 
is that the velocity of a rotating point is $v=\tanh \Omega r$
(for small radius or angular velocities we recovered the situation 
$v=\Omega r$). This adapted coordinate system cover all the Minkowski 
manifold for all angular velocities. Although we will be not able 
to calculate explicity the Bogoliubov coefficients between 
the inertial and the rotating modes we will prove that these coefficients 
are non-zero and in this 
case the answer obtained calculating the Bogoliubov coefficients 
between cartesian and rotating modes and the response function of the 
detector will agree. 

To prove the above assumption, first we have to canonical quantize 
a massless scalar field using this adapted coordinate system
given by eqs.(\ref{7}-\ref{10}).
Making a Taylor expansion for $\cosh\Omega r$ and 
$\tanh\Omega r$ and retaining terms up the first order in $\Omega r$  
the line element becomes
\begin{equation}
{\rm d}s^{2}={\rm d}t^{2}-{\rm d}r^{2}-r^{2}{\rm d}\theta^{2}-
4r\Omega\theta {\rm d}r{\rm d}t-{\rm d}z^{2}. \label{46}
\end{equation}
It is important to stress that the metric given by Eq. (\ref{46}) is an
approximation to the original time dependent metric and the 
solutions that we obtain are approximations to the solutions for 
the modes in the true metric. Note that since we can not solve 
exactly the Klein-Gordon equation without the approximation we 
can not prove the above assumption.
We believe that the approximation in the complete solution is equivalent 
to the exact solution of the Klein-Gordon equation on the 
low velocity limit\footnote{We remark that the Riemann tensor derived using 
the line element given by Eq. (\ref{46}) is null in the considered 
approximation order.}. 

We point out that although we will consider only the case $\Omega r<1$, 
the low-velocity limit of Takeno's transformation does not give 
the ``Galilean" transformation since we have 
\begin{eqnarray}
t&=&t'-\Omega r'^{2}\theta' 
\label{a}
\\
r&=&r', 
\label{b}
\\
\theta&=&\theta'-\Omega t', 
\label{c}
\\
z&=&z'. \label{d}
\end{eqnarray}

Although the angular variables $(\theta,\theta^{'})$ are conected as 
Eq. (\ref{4}) the time-like variables ($t,t^{'}$) has a peculiar 
transformation law. We found such kind of behavior in spinning cosmic
string spacetime, with angular momentum per unit lenght $J$, with line
element:
\begin{equation}
{\rm d}s^{2} = \left( {\rm d}t + 4GJ{\rm d}\theta\right)^{2}-
{\rm d}r^{2} - b^{2}r^{2}{\rm d}\theta^{2}-{\rm d}z^{2}.
\label{V6} 
\end{equation}
$G$ is the Newtonian constant and $b = 1- 4G\mu$, where $\mu$ is the
mass per unit length. This metric is locally flat, as
can immediately be confirmed by changing the coordinates according:
\begin{eqnarray}
\tau &=& t + 4GJ\theta \label{tau}\\
\psi &=& \left(1-4G\mu\right)\theta. \label{psi}
\end{eqnarray}
 
As was discussed by Deser et al \cite{Sousa} to preserve single-valuedness
we must identify time which differ by $8\pi G J$. This gives to the 
spacetime a time helical structure. The difference in both cases is that
in Eq. (\ref{a}) we have a peculiar dependence in $r^2$ instead
$GJ$ in the case of the spinning cosmic string. The possible implications
of this fact is still obscure for us. 

In the low-velocity approximation the metric given by Eq. (\ref{46}) 
is stationary by not static. This means
that although there is a timelike Killing vector field $K$, the spatial
sections putting $t=\mbox{constant}$ are not orthogonal to the time lines 
putting $r,\theta$ and $z$ constants, i.e., the
Killing vector $K$ is not orthogonal to the spatial section. 
This line element describe a physical situation in which world lines
infinitesimally close to a given world line are spatially rotating
with respect to this world line \cite{Frankel}. 

In this simplified case, the Kein-Gordon equation reads
\begin{equation}
\left(\frac{\partial^{2}}{\partial t^{2}}-
\frac{\partial^{2}}{\partial r^{2}}-
\frac{1}{r}\frac{\partial}{\partial r}-
4\Omega\theta\frac{\partial}{\partial t}
-\frac{1}{r^{2}}\frac{\partial^{2}}{\partial\theta^{2}}
-4\Omega\theta r\frac{\partial^{2}}{\partial r\partial t}-
\frac{\partial^{2}}{\partial z^{2}}\right)\varphi(t,r,\theta,z)=0. \label{47}
\end{equation}
The solution can be found using partial separation of variables
\begin{equation}
\varphi(t,r,\theta,z)=T(t)Z(z)f(r,\theta). \label{48}
\end{equation}
Choosing
\begin{equation}
Z(z)=e^{ik_{z}z} \label{49}
\end{equation}
and
\begin{equation}
T(t)=e^{-i\omega t}, \label{50}
\end{equation}
and finally, defining $\omega^{2}=k_{z}^{2}+q^{2}$ we 
obtain the equation for $f(r,\theta)$
\begin{equation}
\left[\frac{\partial^{2}}{\partial r^{2}}+
(\frac{1}{r}-4i\omega\Omega\theta r)\frac{\partial}{\partial r}+
\frac{1}{r^{2}}\frac{\partial^{2}}{\partial\theta^{2}}+
(q^{2}-4i\omega\Omega\theta)\right]f(r,\theta)=0. \label{51}
\end{equation}
The perturbative solution of this equation was derived in the appendix A and
is given by 
\begin{eqnarray}
f_{\mu}(y,\theta) &=& C e^{i\mu\theta}\left[J_{\mu}(y)+ 
l e^{i\lambda\theta}J_{\mu+\lambda}(y)\right]
\nonumber\\
&+ &\frac{l}{2}\int {\rm d}\theta^{'}\int {\rm d}y^{'} 
G(y,\theta ; y^{'},\theta^{'})\theta^{'} \left[y^{'3}J_{\mu-1}(y^{'}) + 2y^{'2}J_{\mu}(y^{'}) -
 y^{'3}J_{\mu+1}(y^{'})\right]. \label{52}
\end{eqnarray}
where $l$, $y$ and $G(y,\theta;y',\theta')$ are also 
defined in the appendix A, and $C$ is a normalization factor.

We must now turn to the question of single valuedness of $f_{\mu}(y,\theta)$.
This situation is very similar to the ($2+1$) dimensional gravity 
\cite{Sousa,Deser}. In our situation we have two different possibilities:
the first is to assume that
$f_{\mu}(y,\theta)$ is a single value function. When $\theta$ increases from $0$ to $2\pi$
for a constant $y$, $t$ jumps by $\Omega r^{2}$ given a time helical structure. It is
possible to show that this solution acquire a phase $e^{i\omega\Omega r^{2}}$.
Using Dirac\rq s arguments and also the Mazur ideas \cite{Dirac} the energy of the modes
must be quantized
$$
\omega = n\left(\frac{2\pi}{\Omega r^{2}}\right),\,\, n=\mbox{integer}.
$$
Note that at principle we can choose that $\Omega$ is quantized to eliminate the
phase problem. With this choice the Bogoliubov coefficients $\beta_{k\nu}$ are zero.
The second one is do not assume that $f_{\mu}(y,\theta)$ is a single value function, and
in this case the Bogoliubov coefficients $\beta_{k\nu}$ are differents from zero
and proportional to $\omega\Omega^{2}r^{2}$. It is important
to realize that if we assume that $f_{\mu}(y,\theta)$ is single value the Minkowski
and the rotating vacuum are the same only in the small
velocity approximation. If we go
further retaing terms of order $\Omega^{2} r^{2}$ 
the Bogoliubov coefficients between the inertial and non inertial modes 
must be different from zero.

Although the spatial part of the solution of eq.(\ref{47}) is 
extremely complicated, there is not ambiguity
in the definition of positive and negative rotating modes 
since the temporal
part is given by eq.(\ref{50}) and the world line of the detector 
is an integral 
curve of the Killing vector $K=\partial/\partial t$ that 
generates a one-parameter group of isometries. 

We have problems to define the 
Hamiltonian in the rotating frame if we work 
with the Takeno coordinate transformation without assume
$\Omega r<1$.
The metric given by eqs.(\ref{11}-\ref{15}) is not 
invariant under time translations.
Usualy the Hamiltonian is defined as
\begin{equation}
H=\int T^{\mu\nu}\xi_{\mu}{\rm d}\sigma_{\nu}\sqrt{-g} \label{53}
\end{equation}
where $\xi_{\mu}$ is a timelike Killing vector field. Since in the 
rotating frame the line element is not stationary it is a complicated 
question how to define $H_{R}$. A possible solution of this 
problem is to use the same idea that we use in expanding universes 
where there is no timelike Killing vector field.  It is 
possible to introduce the definition of particles at each time. This 
procedure introduce the difficult of particle 
creation \cite{Fulling}, and is against the spirit of the 
problem which we are trying to solve.

Before canonical quantize the massless scalar field we will put
in a different way the problem of the rotating detector and the
solution that we obtained for it. Let us accept that the the rotating
frame is a Lorentz frame and the rotating disc 
has a unlimited size. In this case we must use the Takeno\rq s
coordinate system adapted to this rotating frame.
We are faced with a very difficult problem, i.e., to solve the Klein-Gordon
equation without any approximation. We adopted an 
approximation\footnote{Actually we are in the same level as Davies et al solution, since we
considered only the low velocity limit.} and obtained a result totally 
different from Davies et al. The Davies et al rotating vacuum is defined
only in the interior of a cylinder and coincides with the Minkowski
vacuum. Our result is also valid only in a limited region but with
a important physical different result: the rotating vacuum is different
from the Minkowski one. We recognise that the line element given by
Eq. (\ref{46}) put very difficult problems, it is a multivalued metric, etc.
Presumably the correct solution is to solve the full unapproximated
wave equation. In this case the observers would not move along  the 
integral curves of a Killing vector field and the usual quantization
procedure is problematic. Nevertheless it is possible to deal with this
situation. For a careful study of this case see ref. \cite{Magnon}.
If it is possible to implement the canonical quantization procedure
in this non stationary situation the resulting  rotating vacuum would be well described.  In the
absence of such solution we continue to use the low velocity approximation.

Going back to the low-limit velocity case, we have that
the line element is stationary and there is no ambiguity to define 
the rotating vacuum
$|0,R>$ is such a way that:
\begin{equation}
b_{q\mu k_{z}}|0,R>=0. \qquad\forall\qquad  q,\mu,k_{z},\label{54}
\end{equation}
where 
\begin{equation}
\varphi(t,r,\theta,z)=
\sum_{\mu}\int q{\rm d}q{\rm d}k_{z}\left[b_{q\mu k_{z}}v_{q\mu k_{z}}(t,r,\theta,z)+
b_{q\mu k_{z}}^{\dag}v_{q\mu k_{z}}^{*}(t,r,\theta,z)\right]. \label{55}
\end{equation}
By sake of simplicity let use the following notation:
\begin{equation}
\varphi(t,r,\theta,z)
=\sum_{\nu}b_{\nu}v_{\nu}(t,r,\theta,z)+
b_{\nu}^{\dag}v_{\nu}^{*}(t,r,\theta,z), \label{56}
\end{equation}
where $\nu\equiv\{q,\mu,k_{z}\}$ is a collective index.

It is straighforward to show that
the Minkowski vacuum can be expressed as a many rotating-particles 
state. By comparing the expansion of the field operator using the 
inertial modes and the rotating modes it is possible 
to obtain the expression comparing both vacua, i.e $|0,M>$ and $|0,R>$:
\begin{equation}
|0,M>=
e^{\frac{i}{2}\sum_{\mu,\nu}b^{\dag}(\mu)V(\mu,\nu)b^{\dag}(\nu)}|0,R> \label{57}
\end{equation}
where
\begin{equation}
V(\mu,\nu)=i\sum_{k}\beta^{*}_{\mu k}\alpha^{-1}_{k\nu}, \label{58}
\end{equation}
and the Bogoliubov coefficients  are given by 
$\beta_{\nu k}=-(v_{\nu},\psi^{*}_{k})$ and 
$\alpha_{\nu k}=(v_{\nu},\psi_{k})$ .
It is clear that the number of rotating particles in a specific mode 
in the Minkowski vacuum is given by
\begin{equation}
<0,M|N_{R}(\nu)|0,M>=\sum_{k}|\beta_{k\nu}|^{2}. \label{59}
\end{equation}
Let us choose the hypersurface $t'=\mbox{constant}$ to find 
the Bogoliubov coeficients, i.e.,
\begin{equation}
\beta_{\nu k}=i\int_{0}^{2\pi}{\rm d}\theta'\int^{\infty}_{-\infty}{\rm d}z
\int^{\infty}_{0}r{\rm d}r\left\{v_{\nu}(x')\left[\partial_{t'}\psi_{k}(x')\right]
-\left[\partial_{t'}v_{\nu}(x')\right]\psi_{k}(x')\right\}. \label{60}
\end{equation}
The Bogoliubov coefficients $\beta_{\nu k}$ must be 
non-zero since the positive and negative frequency rotating modes are 
mixture between positive and negative inertial modes.
The important conclusion from the above arguments is that the 
Minkowski and this rotating vacuum are not the same.
Now we will show that the expectation value of the number operator 
of rotating particles is proportional to the response function, recovering
the old idea that rate of excitation is proportional to the 
number of particles (in the mode of interest) in the state of the field. 
Note that the rate given by eq.(\ref{43}) can be written as:
\begin{equation}
R(E,\Delta T,\xi,\xi')=
R_{I}(E,\Delta T,\xi,\xi') +
R_{II}(E,\Delta T,\xi,\xi') \label{61}
\end{equation}
where
\begin{eqnarray}
R_{I}(E,\Delta T,\xi,\xi')=
\int^{\Delta T}_{-\Delta T}{\rm d}\zeta
e^{iE\zeta}
\left(\left<0,M\right|\varphi^{(-)}(\eta^{'},\xi^{'})
\varphi^{(+)}(\eta,\xi)\left|0,M\right>\right.
\nonumber \\ 
+ \left.\left<0,M\right|\varphi^{(-)}(\eta,\xi)
\varphi^{(+)}(\eta',\xi')\left|0,M\right>\right) \label{62}
\end{eqnarray}
and
\begin{equation}
R_{II}(E,\Delta T,\xi,\xi')=
\int^{\Delta T}_{-\Delta T}d\zeta
e^{iE\zeta}
\left<0,M\right|[\varphi^{+}(\eta^{'},\xi^{'}),\varphi^{-}
(\eta,\xi)]\left|0,M\right>. \label{63}
\end{equation}
Let us investigate the contribution given by Eqs. (\ref{62}) and  (\ref{63}) to the
rate assuming the detector at rest in the rotating frame. Substituting Eqs. (\ref{49}),
(\ref{50}) and (\ref{52}) in Eq. (\ref{63}) we have for $R_{II}(E,\Delta T,\xi)$
\begin{equation}
R_{II}(E,\Delta T,\xi)= 
\sum_{\mu\nu}\delta_{\mu\nu}
|Z|^{2}f_{\mu}(\xi)f^{*}_{\nu}(\xi)
\int^{\Delta T}_{-\Delta T}{\rm d}\zeta e^{iE\zeta}
\left[e^{-i\omega_{\mu}(\eta-\frac{\omega_{\nu}}{\omega_{\mu}}\eta')}+
e^{i\omega_{\nu}(\eta-\frac{\omega_{\mu}}{\omega_{\nu}}\eta')}\right]. \label{64}
\end{equation}
It is not difficult to perform the $\zeta$ integral and obtain
\begin{equation}
R_{II}(E,\Delta T,\xi) = \Delta T \sum_{\mu}|Z|^{2}
|f_{\mu}(\xi)|^{2}
\left\{\frac{\sin\left[(E-\omega_{\mu})\Delta T\right]}{(E-\omega_{\mu})\Delta T}
+\frac{\sin\left[(E+\omega_{\mu})\Delta T\right]}{(E+\omega_{\mu})\Delta T}\right\}. \label{65}
\end{equation}
In the asymptotic limit we have
\begin{equation}
\lim_{\Delta T \rightarrow\infty}R_{II}(E,\Delta T,\xi) =\sum_{\mu}  |Z|^{2}|f_{\mu}(\xi)|^{2}(\delta(E-\omega_{\mu})+\delta(E+\omega_{\mu})) . \label{66}
\end{equation}
Note that in $R_{II}(E,\Delta T,\xi)$ we have two processes: absorption and 
emission processes of rotating particles. 
This contribution to the rate does not depend 
on the number of rotating  particles in the Minkowski vacuum. 
The $R_{II}(E,\Delta T,\xi)$ contribution is independent of the 
particle content of the state of the field. This is the reason that Glauber and others
disregarded the $R_{II}(E,\Delta T,\xi)$ term to the rate. It is important to have in
mind although $R_{II}(E,\Delta T,\xi)$ is not related with the particle content of
the state, it is responsible for the spontaneous decay process.
To have a better understanding of the meaning of $R_{II}(E,\Delta T,\xi)$ let us 
repeat the calculations preparing the field in the rotating vacuum i.e.
\begin{equation}
R_{II}(E,\Delta T,\xi,\xi')=
\int^{\Delta T}_{-\Delta T}d\zeta
e^{iE\zeta}
\left<0,R\right|[\varphi^{+}(\eta^{'},\xi^{'}),\varphi^{-}
(\eta,\xi)]\left|0,R\right>. \label{nova2}
\end{equation}
In the case of spontaneous 
excitation $(E>0)$ the 
asymptotic limit for eq.(\ref{nova2}) gives zero 
(stability of the detector\rq s ground state). For the rotating detector interacting
with the field in the Minkowski vacuum, we will have this term plus a term proportional
to $\Omega^2 r^2$, i.e., for the case of spontaneous decay we have 
\begin{equation}
\lim_{\Delta T\rightarrow \infty}R_{II}(E,\Delta T)=
\frac{|E|}{2\pi}(1-16\Omega^{2}r^{2}) \label{nova1}
\end{equation}
The probability of decay per unit time decrease with the square of the distance
from the origin. Since we are using the approximation $\Omega^2 r^2 < 1$, Eq. (\ref{nova2})
never becomes negative.
Let us analize the contribution given by $R_{I}(E,\Delta T,\xi)$.
It is not difficult to show that the term between the parentesis 
in eq.(\ref{62}) gives
\begin{eqnarray}
\left<0,M\right|\varphi^{(-)}(\eta^{'},\xi^{'})
\varphi^{(+)}(\eta,\xi)\left|0,M\right>
 + \left<0,M\right|\varphi^{(-)}(\eta,\xi)
\varphi^{(+)}(\eta',\xi')\left|0,M\right>=\nonumber\\
\sum_{\mu\nu}\sum_{k}\beta^{*}_{\nu k}\beta_{k\mu}
\left[v_{\nu}(\eta',\xi')v^{*}_{\mu}(\eta,\xi)+
v_{\nu}(\eta,\xi)v^{*}_{\mu}(\eta',\xi')\right]. \label{67} 
\end{eqnarray}
Consequently the $R_{I}(E,\Delta T,\xi)$ contribution to the rate of 
transition for the detector at rest in the rotating frame is:
\begin{equation}
R_{I}(E,\Delta T,\xi)= 
\sum_{\mu\nu}\sum_{k}\beta^{*}_{\mu k}\beta_{k\nu}
|Z|^{2}f_{\mu}(\xi)f^{*}_{\nu}(\xi)
\int^{\Delta T}_{-\Delta T}d\zeta e^{iE\zeta}
\left[e^{-i\omega_{\mu}(\eta-\frac{\omega_{\nu}}{\omega_{\mu}}\eta')}+
e^{i\omega_{\nu}(\eta-\frac{\omega_{\mu}}{\omega_{\nu}}\eta')}\right]. \label{68}
\end{equation}
In this expression there are two different contributions: the non-diagonal and the 
diagonal terms ($\mu=\nu$). 
Let us analize the contribution to the rate from the diagonal terms given by
\begin{equation}
R_{I}(E,\Delta T,\xi)= 
\sum_{\mu}\sum_{k}\beta^{*}_{\mu k}\beta_{k\mu}|Z|^{2}f_{\mu}(\xi)f^{*}_{\mu}(\xi)
\int^{\Delta T}_{-\Delta T}{\rm d}\zeta e^{iE\zeta}
\left[e^{-i\omega_{\mu}(\eta-\eta')}+
e^{i\omega_{\mu}(\eta-\eta')}\right]. \label{69}
\end{equation}
It is not difficult to perform the $\zeta$ integral and using eq.(\ref{59}) we obtain
\begin{eqnarray}
R_{I}(E,\Delta T,\xi) = \Delta T \sum_{\mu}|Z|^{2}
|f_{\mu}(\xi)|^{2}\left<0,M|N_{R}(\mu)|0,M\right>\nonumber\\
\times\left\{\frac{\sin\left[(E-\omega_{\mu})\Delta T\right]}{(E-\omega_{\mu})\Delta T}
+\frac{\sin\left[(E+\omega_{\mu})\Delta T\right]}{(E+\omega_{\mu})\Delta T}\right\} \label{70}
\end{eqnarray}
In the asymptotic limit the rate of transition becomes
\begin{equation}
\lim_{\Delta T \rightarrow\infty}R_{I}(E,\Delta T,\xi) =  \sum_{\mu}|Z|^{2}|f_{\mu}(\xi)|^{2}\left<0,M|N_{R}(\mu)|0,M\right>
(\delta(E-\omega_{\mu})+\delta(E+\omega_{\mu})) . \label{71}
\end{equation}
We have two different processes in the above expression: absorption of
rotating particles $(E>0)$ and induced emission of rotating particles 
$(E<0)$. In both cases
 the rate of transition will be proportional to the number 
of rotating particles with energy $E$ in the  Minkowski vacuum 
multiplied by the square of the 
``wavefunction" ($|Z||f(r,\theta)|$) in the world line of the detector. 
This situation is exactly the case of a detector in a thermal bath where the 
rate of absorption of the particles of the bath is equal to the rate of induced emission.
Note that we still have to the rate of absorption and emission processes 
the contribution to the non-diagonal terms.   

Bell and Leinaas studied the depolarization problem in accelerators trying 
to use the idea of a Unruh-Davies effect. 
The electron in a accelerated 
ring is a magnetic version of the monopole detector, since there 
is a linear coupling between the magnetic field $B$ and the magnetic moment 
of the electron. 
To see this result let us define the invariant operator 
\begin{equation}
H=\frac{e}{2m^{2}}F^{*}_{\mu\nu}p^{\mu}s^{\nu} \label{72}
\end{equation}
where $m$ is the electron mass, $F^{*}_{\mu\nu}=\frac{1}{2}
\epsilon_{\mu\nu\rho\sigma}F^{\rho\sigma}$ and $s^{\nu}$ is the 
four vector spin operator. In the frame in which the electron 
is at rest the operator $H$ describes the interaction between the 
spin magnetic moment of the electron with the magnetic field,
\begin{equation}
H=-\vec{\mu}.\vec{B}. \label{73}
\end{equation}
To understand the depolarization problem let 
us suppose an ensemble of detectors in equilibrium 
with a thermal bath. The probability to find the  detector in the 
state $|i>$ is:
\begin{equation}
P_{i}=\frac{e^{-\beta\omega_{i}}}{Z} \label{74}
\end{equation}
or 
\begin{equation}
\frac{P_{e}}{P_{g}}=e^{-\beta E} \label{75}
\end{equation}
defining the occupation number $N(e)$ and $N(g)$ we have 
\begin{equation}
N(e)=N(g)e^{-\beta E}. \label{76}
\end{equation}
Since the electron in a accelerator is a magnetic version of the 
monopole detector, in the equilibrium the rate between spin up and spin
down will be given by the above equation. Thus if we introduce 
a complete unpolarized electron beam, it will suffer a polarization until 
the equilibrium is reached. The asymptotic rate of spin flip will be
proportional to the asymptotic limit of 
the rate $R_{\beta}(E,\Delta T)$ i.e.,
\begin{equation}
\lim_{\Delta T\rightarrow \infty}R_{\beta}(E,\Delta T)=
\frac{|E|}{2\pi}\left[\Theta(-E)\left(1+\frac{1}{e^{\beta |E|}-1}\right)+
\Theta(E)\frac{1}{e^{\beta E}-1}\right]. \label{77}
\end{equation}

Note that although the situation is similar 
to the Rindler's case where the detector goes to excited state 
by absorption of Rindler's particles (the Minkowski vacuum is a 
thermal state of Rindler's particles), there is a fundamental difference.
In the Rindler's case there is an past and future horizont.
Part of information which would have an inertial observer is inaccessible 
for accelerated observers. Although the Minkowski vacuum $|0,M>$ 
is a pure state, for accelerated observers it must be described 
by a statistical operator. This is the origin of the 
thermal distribution of particles. 
As was noted by Bell and Leinaas 
in the case of circular motion the measurements of the polarization 
{\em does not agree} with the calculations if we 
interpret the polarization by 
thermal effects. In our approach, depolarization is 
related with the fact 
that the Minkowski vacuum is a many particle state 
of rotating particles with a non thermal spectrum.  
Let us try to improve this ideas using Einstein's 
arguments \cite{Einstein}.
All calculations will be held in the rotating frame.  
Suppose that
the probability to find the detector in the state $|i>$ is given by 
\begin{equation}
P(\omega_{i})=\frac{f(\omega_{i})}{Z} \label{78}
\end{equation}
where the partition function $Z$ is given by
\begin{equation}
Z=\sum^{2}_{i=1}f(\omega_{i}). \label{79}
\end{equation} 
Still following Einstein's arguments we have three different processes:
absortion of rotating particles, induced emission 
and spontaneous emission (stimulated emission by 
the $|0,R>$ vacuum fluctuations) of rotating particles.
Defining the rate of spontaneous decay by $A_{2\rightarrow 1}(E,\Delta T)$ 
we have 
\begin{equation}
{\rm d}W_{2\rightarrow 1}(E,\Delta T)= A_{2\rightarrow 1}(E,\Delta T){\rm d}t \label{80}
\end{equation} 
For the induced emission $R_{2\rightarrow 1}(E,\Delta T)$ we have
\begin{equation}
{\rm d}W_{2\rightarrow 1}(E,\Delta T)= R_{2\rightarrow 1}(E,\Delta T){\rm d}t, \label{81}
\end{equation}
and finally for the rate of absorption 
$R_{1\rightarrow 2}(E,\Delta T)$   we have 
\begin{equation}
{\rm d}W_{1\rightarrow 2}(E,\Delta T)= R_{1\rightarrow 2}(E,\Delta T){\rm d}t. \label{82}
\end{equation}
In the {\it equilibrium situation} between an ensemble of 
rotating detectors and the 
scalar field in the Minkowski vacuum (asymptotic limit) we have
\begin{equation}
f(\omega_{1})\rho(E) R_{1\rightarrow 2}(E)
=f(\omega_{2})\left[\rho(E) R_{2\rightarrow 1}(E)+
A_{2\rightarrow 1}(E)\right] \label{83}
\end{equation}
where $\rho$ is the number of rotating particle in the mode $E$ in the 
Minkowski vacuum i.e. 
\begin{equation}
\rho(E)=<0,M|N_{R}(E)|0,M>. \label{84}
\end{equation}
Although the spectrum of the rotating particles in the Minkowski vacuum is
not known, at the equilibrium we have 
$R_{1\rightarrow 2}(E)=R_{2\rightarrow 1}(E)$. In the 
equilibrium situation the this hipotesis must hold. Note that this is 
not in principle fundamental for our conclusions. 
A straightforward calculations gives
\begin{equation}
\rho(E)=\frac{A_{2\rightarrow 1}(E)}{R_{1\rightarrow 2}(E)}
\frac{1}{\frac{f(\omega_{1})}{f(\omega_{2})}-1} \label{85}
\end{equation}
The knowledge of the Bogoliubov coefficients $\beta_{k\mu}$ 
give us both $\rho(E)$ and $R_{2\rightarrow 1}(E)$. 
A second step in our 
analysis is to use the result that $A_{2\rightarrow 1}(E)$ is exactly the 
rate of spontaneous decay of a inertial detector interacting with 
the field in the Minkowski vacuum. Thus we have
\begin{equation}
\frac{f(\omega_{1})}{f(\omega_{2})}=
\left(\frac{E R_{1\rightarrow 2}^{-1}(E)}{<0,M|N_{R}(E)|0,M>}-1\right). \label{86}
\end{equation}
This result show us the connection between the the rate between up and 
down spins as a function of the mean life of the excited state and $\rho(E)$
after the equilibrium situation is reached.

We still have to answer some questions. Where does the energy of 
excitation come from if we analyse the process from 
the point of view 
of the inertial observer? The non-inertial observer 
does not meet any difficulty.
At some initial time we 
prepare the detector in the ground state and the field 
in the Minkowski 
vacuum. Since the Minkowski vacuum is a many rotating-particles state 
the detector goes to excited state absorbing a 
positive energy particle. 
For large time intervals energy conservation holds. For the point of 
view of the inertial observer the field is in the 
empty state. How is 
possible the excitation? A natural answer is to say that it is necessary an 
external accelerating agency to suplly energy to 
keep the detector 
in the rotating world-line. It is possible to show that the detector 
goes to excited state with the emission of a 
Minkowski particle.
In the next section we will perform the second 
quantization of the detector Hamiltonian to analyse the 
absorption and emission 
processes from the inertial point of view.

\section{Second Quantization of the Total Hamiltonian and Polarization 
Effects on Electrons and Positrons in Storage Rings}

In this section we will prove that the process: absorption (emission) of 
positive energy rotating particle with excitation (decay) 
of the detector (from the 
non-inertial point of view) is interpreted as a emission of a 
Minkowski particle with excitation (decay) of the detector from the 
inertial point of view. This simple result express the fact that 
electrons (positrons) experience a gradual polarization 
orbiting in a storage ring.
This mechanism lead to the emission of spin-flip synchronon radiation
\cite{Jackson}. It is important to stress that the amount of spin-flip 
radiation is extremely small compared with the non-flip 
radiation. An open 
question is why the polarization is not complete after the system reach 
the equilibrium? We will try to answer this question applying the ideas 
developed by us in the preceding sections. 
Of course again we have a oversimplified description 
of the phenomenon. Before start the second quantization of the 
detector and interaction Hamiltonian let us remember the fundamental
results of the preceeding section (we will use a different notation in 
this section). 

In Minkowski space time it is possible 
to quantize a massless scalar field using the 
cartesian coordinate adapted to inertial observers. Thus the 
scalar field can be expanded using an orthonormal set of modes 
\begin{equation}
\varphi(x)=\sum_{i}a_{i}u_{i}(x)+a^{\dag}_{i}u^{*}_{i}(x) \label{87}
\end{equation}
where 
\begin{equation}
a_{i}|0,M>=0 \qquad\qquad \forall\, i. \label{88}
\end{equation}
A rotating observer can also second quantize the scalar field and 
this quantization is unitarily non-equivalent to the quantization 
implemented by inertial observers. Thus 
the scalar field can be expanded using a 
second set of orthonormal modes
\begin{equation}
\varphi(x)=\sum_{j}b_{j}v_{j}(x)+b^{\dag}_{j}v^{*}_{j}(x) \label{89}
\end{equation}
where 
\begin{equation}
b_{j}|0,R>=0 \qquad\qquad \forall\, j. \label{90}
\end{equation}
As both sets are complete, the non-inertial modes can be expanded 
in terms of the inertial ones, i.e.
\begin{equation}
v_{j}(x)=\sum_{i}\alpha_{ji}u_{i}(x)+\beta_{ji}u^{*}_{i}(x) \label{91}
\end{equation}
or
\begin{equation}
u_{i}(x)=\sum_{j}\alpha^{*}_{ji}v_{j}(x)-\beta_{ji}v^{*}_{j}(x). \label{92}
\end{equation}
Using these equations and the orthonormality of the modes it is 
possible to write the annihilation and creation operators of inertial 
particles in the mode $i$ as a linear combination 
of non-inertial creation and annihilation operators \cite{Birrel}, i.e.
\begin{equation}
a_{i}=\sum_{j}\alpha_{ji}b_{j}+\beta^{*}_{ji}b_{j}^{\dag} \label{93}
\end{equation}
or
\begin{equation}
b_{j}=\sum_{i}\alpha^{*}_{ji}a_{i}-\beta^{*}_{ji}a^{\dag}_{i}. \label{94}
\end{equation}
Let us use the notation introduced in section 3, i.e. $|g>=|1>$ and 
$|e>=|2>$. Thus we have
\begin{equation}
H_{D}|i>=\omega_{i}|i> \qquad\qquad i=1,2. \label{95}
\end{equation}
Using the above equation and the orthonormality of the energy 
eigenstates of the detector Hamiltonian, we can write
\begin{equation}
H_{D}=\sum_{i=1}^{2}\omega_{i}|i><i|. \label{96}
\end{equation}

To second quantize the detector Hamiltonian  
we have to introduce the Dicke operators \cite{Dicke}
\begin{equation}
S^{+}=|2><1|, \label{97}
\end{equation}
\begin{equation}
S^{-}=|1><2|, \label{98}
\end{equation}
and finally 
\begin{equation}
S_{z}=\frac{1}{2}(|2><2|-|1><1|). \label{99}
\end{equation}
In the case of $n$ eigenstates of the (atom) detector Hamiltonian 
we have to work with the atomic operators, i.e. the multilevel 
generalization of the Dicke spin operators for the two level system.
The detector Hamiltonian in the two level case can be written as
\begin{equation}
H_{D}=ES_{z}+\frac{1}{2}(\omega_{1}+\omega_{2}). \label{100}
\end{equation}
The operators $S^{+}$, $S^{-}$ and $S_{z}$ satisfy the angular 
momentum commutation relations corresponding to spin $1/2$ value, i.e.
\begin{eqnarray}
\left[S^{+},S^{-}\right]&=&2S_{z}, \label{101} \\
\left[S_{z},S^{+}\right]&=&S^{+}, \label{102} \\
\left[S_{z},S^{-}\right]&=&-S^{-}. \label{103}
\end{eqnarray}
It is clear that $S^{+}$ and $S^{-}$ are respectivelly raising  
and lowering operators of the detector states ($S^{+}|1>=|2>,S^{+}|2>=0,
S^{-}|2>=|1>,S^{-}|1>=0$).
The interaction Hamiltonian given by eq.(\ref{25}) can be written as 
\begin{equation}
H_{int}=\lambda[m_{21}S^{+}+m_{12}S^{-}+
S_{z}(m_{22}-m_{11})]\varphi(x), \label{104}
\end{equation}
where
\begin{equation}
<i|m(0)|j>=m_{ij},
\label{105}
\end{equation}
and $\varphi(x)$ must be evaluated in the world line of the detector. 
We should simplify the discussion choosing $m_{11}=m_{22}$. 
As we will see
the part of the interaction hamiltonian with the $S_{z}$ operator is 
responsible for the non-flip synchroton radiation.
Substituting eq.(\ref{89}) in eq.(\ref{104}) we see that there are different 
processes with absorption or emission of rotating particles 
with excitation or decay of the detector. It is possible to show 
that some of these processes are virtual, and only processes 
with energy conservation survive in the asymptotic limit, i.e.,
excitation of the detector with absorption of a rotating particle
(process involving $b_{j}S^{+}$) and decay of the 
detector with emission of a 
rotating particle (process involving $b^{\dag}_{j}S^{-}$).

The first process is generated by the following operators:
\begin{equation}
m_{12}\sum_{j}v_{j}(x)b_{j}S^{+}. \label{106}
\end{equation}
Substituting eq.(\ref{91}) and eq.(\ref{94}) in eq.(\ref{106}) it is
clear that the above 
process of absorption of a rotating particle in the mode $j$ is
the following:
\begin{equation}
\sum_{ijk}\left[\beta^{*}_{ji}\alpha_{jk}u_{k}(x)+
\beta_{ji}^{*}\beta_{jk}u_{k}^{*}(x)\right]a_{i}^{\dag}S^{+}. \label{107} 
\end{equation} 
Therefore this process for the inertial observer is an 
excitation of the detector with creation of Minkowski particles. 

The second process is generated by the following operators:
\begin{equation}
m_{21}\sum_{j}v^{*}_{j}(x)b^{\dag}_{j}S^{-}. \label{108}
\end{equation}
Substituting eq.(\ref{91}) and eq.(\ref{94}) in eq.(\ref{108})
we see that the above  
process of emission of a rotating particle in the mode $j$ is
the following:
\begin{equation}
\sum_{ijk}\left[\alpha_{ij}\alpha^{*}_{jk}u^{*}_{k}(x)+
\alpha_{ij}\beta^{*}_{jk}u_{k}(x)\right]a_{i}^{\dag}S^{-}. \label{109}
\end{equation} 
Therefore this process for the inertial observer is a
decay of the detector with creation of Minkowski particles.

Now we are able to understand the problem of the synchroton 
radiation. In the emision of synchroton radiation by 
electrons moving along a circular orbit, there are two kinds 
of processes: the first is the emission of photons 
without spin flip of the electron and the second is emission with 
spin flip. We will restrict our discussion to the second case. 
To make a parallel with the detector's problem we have to assume 
that the electron trajectory is ``classical" (there is no 
fluctuation of the electron path) or even after the 
photon emission there is no recoil (as was stressed by Bell 
and Leinaas, the 
results does not depend on position fluctuations of 
the electron trajectory).
There are two differents 
results in the literature depending on the value of the Land\'e-g factor 
of the electron. Jackson 
showed that the rate of transition from an initial 
state with the spin of the electron directed along the magnetic 
field (high energy state) to a state with the electron spin 
in opposite to the magnetic field (lower energy state) is lower 
than the opposite situation if the Land\'e-g factor of the
electron obeys $0<g<1.2$. It is important to stress that 
the situation
is opposite of the naive description where polarization arises from 
spontaneous emission as the spin move 
from its ``upper" (high energy state) 
to its ``lower" (low energy state) in the magnetic field. For the
case where $1.2<g<2$ Jackson and also Sokolov et al \cite{Sok} 
obtained that 
after the photon emission 
the electron spin will tends to orient themselves in 
opposite to the magnetic field (going to the lower energy state).
Of course, positrons spins will have an oposite behavior.
These both results are consistent with our interpretation
that absorption (emission) of a rotating particle with excitation 
(decay) of the detector in the non-inertial frame is interpreted
as emission of a Minkowski particle 
with excitation (decay) of the detector in the inertial frame. 

To find the degree of polarization before the equilibrium situation 
is achieved let us define the occupation number of electrons 
with spins directed in oposition to the magnetic 
field (lower energy state) by 
$N_{1}$, and $N_{2}$ is the number of electrons with spins directed 
to the magnetic field. Of course we have
$N_{1}(t)+N_{2}(t)=N$, where $N =\mbox{constant}$ is the 
total numbers of electrons in the ring. 
We will do all the calculations in the rotating frame.
The degree of polarization of an ensemble of electrons 
in the beam is defined as 
\begin{equation}
P(t)=\frac{N_{1}(t)-N_{2}(t)}{N_{1}(t)+N_{2}(t)}. \label{110}
\end{equation}
The equation of the evolution of $N_{1}$ and $N_{2}$ are given by 
\begin{equation}
\frac{dN_{1}}{dt}=N_{2}\left[\rho(E) R_{2\rightarrow 1}(E,\Delta T)
+ A_{2\rightarrow 1}(E,\Delta T)\right] - 
N_{1}\left[\rho(E) R_{1\rightarrow 2}(E,\Delta T)\right] \label{111}
\end{equation}
and
\begin{equation}
\frac{dN_{2}}{dt}= N_{1}\left[\rho(E) R_{1\rightarrow 2}(E,\Delta T)\right]-
N_{2}\left[\rho(E)R_{2\rightarrow 1}(E,\Delta T)+
A_{2\rightarrow 1}(E,\Delta T)\right] \label{112}
\end{equation}
Let us avoid the difficult to find $R_{1\rightarrow 2}(E,\Delta T))$ and 
$R_{2\rightarrow 1}(E,\Delta T)$ and using the following 
approximation, i.e., 
\begin{equation}
\rho(E)R_{2\rightarrow 1}(E,\Delta T)+ A_{2\rightarrow 1}(E,\Delta T)
=\sigma_{21}=\mbox{constant} \label{113}
\end{equation}
and
\begin{equation}
\rho(E)R_{1\rightarrow 2}(E,\Delta T)=\sigma_{12}=\mbox{constant}. \label{114}
\end{equation}
We are choosing the asymptotic limit (see eq.(\ref{66}) and eq.(\ref{71})).
Then, starting from a situation where there is no polarization, i.e., 
$P(t=0)=0$ it is possible 
to find the polarization until the equilibrium situation is
achieved. It is necessary only to integrate the above equations.  
A straightforward calculation gives
\begin{equation}
N_{1}(t)=\frac{N}{2}\left(\frac{\sigma_{12}-\sigma_{21}}{\sigma_{12}
+\sigma_{21}}\right) e^{-(\sigma_{12}+\sigma_{21})t} + 
N\left(\frac{\sigma_{21}}{\sigma_{12}+\sigma_{21}}\right) \label{115} 
\end{equation}
and
\begin{equation}
N_{2}(t)=-\frac{N}{2}\left(\frac{\sigma_{12}-
\sigma_{21}}{\sigma_{12}+\sigma_{21}}\right) e^{-(\sigma_{12}+\sigma_{21})t}
+ N\left(\frac{\sigma_{12}}{\sigma_{12}+\sigma_{21}}\right). \label{116}
\end{equation}
The degree of polarization of the beam is
\begin{equation}
P(t)=\left(\frac{\sigma_{21}-\sigma_{12}}{\sigma_{12}+
\sigma_{21}}\right)\left(1-e^{-(\sigma_{12}+\sigma_{21})t}\right). \label{117}
\end{equation}

We obtained that if $R_{1\rightarrow 2}(E,\Delta T)$, $R_{2\rightarrow 1}(E,\Delta T)$
and $A_{2\rightarrow 1}(E,\Delta T)$ are independent of time the asymptotic degree 
of polarization is constant i.e., 
\begin{equation}
\lim_{t\rightarrow\infty}P(t)=\left(\frac{\sigma_{21}-\sigma_{12}}
{\sigma_{12}+\sigma_{21}}\right). \label{118}
\end{equation}
Experimental results 
show us  a not complete polarization. Why there is residual depolarization?   
This is the puzzle 
stressed by Jackson \cite{Jackson} and also Bell and Leinas \cite{Bell}.
From the former equation it is easy to see that the polarization 
can not be complete. The process absorption of a rotating particle with excitation
of the detector has always non null probability.
In the asymptotic limit we have that if
\begin{equation}
R_{21} + A_{21} > 3R_{12}, \label{119}
\end{equation}
the lower energy state is prefered ($1.2<g<2$, for 
the Land\'e-g factor), and
if
\begin{equation}
R_{21} + A_{21} < 3R_{12}, \label{120}
\end{equation}
the higher energy state is prefered ($0<g<1.2$ for the Land\'e-g factor). 

We remark that the results that the polarization can not be complete
was obtained in a very crude approximation where 
the rates $R_{1\rightarrow 2}(E,\Delta T)$, 
$R_{2\rightarrow 1}(E,\Delta T)$ and $A_{2\rightarrow 1}(E,\Delta T)$
does not depend on time [see eq.(\ref{65}) and eq.(\ref{70})].  
A more realistic result
can be obtained assuming that this rates does depend on time.
Defining $n_{1}=N_{1}/N$ and $n_{2}=N_{2}/N$ and also 
\begin{equation}
\rho(E)R_{2\rightarrow 1}(E,\Delta T)+A_{2\rightarrow 1}(E,\Delta T) \label{121}
=\sigma_{21}(t)
\end{equation}
and
\begin{equation}
\rho(E)R_{1\rightarrow 2}(E,\Delta T)=\sigma_{12}(t) \label{122}
\end{equation}
we obtain the following equations:
\begin{equation}
n_{1}(t)+n_{2}(t)=1 \label{123}
\end{equation}
and
\begin{equation}
\frac{{\rm d}n_{1}(t)}{{\rm d}t}+n_{1}(t)\left[\sigma_{12}(t)+
\sigma_{21}(t)\right]=\sigma_{21}(t) \label{124}
\end{equation}
Let us consider the homogeneous linear equation:
\begin{equation}
\frac{{\rm d}n_{1}^{(0)}(t)}{{\rm d}t}=n_{1}^{(0)}(t)\left[\sigma_{12}(t)+
\sigma_{21}(t)\right]=0 \label{125}
\end{equation}
A general solution is
\begin{equation}
n^{(0)}_{1}(t)=C_{1}e^{-\int^{t}\left[\sigma_{12}(t')+
\sigma_{21}(t')\right]{\rm d}t'}. \label{126}
\end{equation}
Now let us substitute in the non-homogeneous equation the expression
\begin{equation}
n_{1}(t)=v(t)e^{-\int^{t}\left[\sigma_{12}(t')+
\sigma_{21}(t')\right]{\rm d}t'}. \label{127}
\end{equation}
The equation for $v(t)$ becomes
\begin{equation}
\frac{{\rm d}v(t)}{{\rm d}t}e^{-\int_{0}^{t}\left[\sigma_{12}(t')+
\sigma_{21}(t')\right]{\rm d}t'}=\sigma_{21}(t) \label{128}
\end{equation}
consequently we have
\begin{equation}
v(t)=C_{2}+\int^{t}{\rm d}t'\sigma_{21}(t')
e^{\int^{t'}\left[\sigma_{12}(t'')+\sigma_{21}(t'')\right]{\rm d}t''}. \label{129}
\end{equation}
The general solution that we are looking for 
involves two quadratures and it is given by
\begin{equation}
n_{1}(t)=C_{2}e^{-\int^{t}\left[\sigma_{12}(t')+
\sigma_{21}(t')\right]{\rm d}t'}+
e^{-\int^{t}\left[\sigma_{12}(t')+\sigma_{21}(t')\right]{\rm d}t'}
\int^{t}{\rm d}t'\sigma_{21}(t')
e^{\int^{t'}\left[\sigma_{12}(t'')+\sigma_{21}(t'')\right]{\rm d}t''}. \label{130}
\end{equation}
With the values of $R_{2\rightarrow 1}(E,\Delta T)$, 
$R_{1\rightarrow 2}(E,\Delta T)$ and $A_{2\rightarrow 1}(E,\Delta T)$
(given by eqs.(\ref{65}) and (\ref{70})) it is possible to find the degree of polarization.

We would like to point out that 
there is a different approach to study these problems.
As it has been pointed out by Milonni and 
Smith \cite{Milonni} and Ackerhalt,
Knight and Eberly \cite{Knight}, it is possible  
to study
radiative processes without using perturbation theory, 
but using the 
Heisenberg equations of motion. In this approach it is 
possible to
obtain non-perturbative expressions for radiative processes 
where the
radiation reaction appears in a very simple way: 
the part of the field 
due to the atom (detector) that drives the Dicke 
operators \cite{Dicke}.
In this approach it is possible to identify the role 
of radiation reaction
and vacuum fluctuations in spontaneous emission. We 
would like to stress 
the fact that the contribution of vacuum fluctuations 
and radiation 
reaction can be chosen arbitrarily, depending on the 
order of the Dicke 
and field operators. As it was discussed by Dalibard, 
Dupont-Roc and 
Cohen-Tannoudji \cite{Cohen}, there is a preferred ordering 
in such a
way that the vacuum fluctuations and radiation reaction 
contribute
equally to the spontaneous emission process. More 
recently this approach
was developed by Audretsch and Muller, and also 
Audretsch, Muller
and Holzmann \cite{Audretsch 1} to study the 
Unruh-Davies 
effect.
These authors constructed the following picture of the 
Unruh-Davies effect.
The effect of vacuum fluctuations is changed by the 
acceleration, 
although the contribution of radiation reaction is unaltered.
Due to the modified vacuum fluctuation contribution, 
transition to an
excited state becomes possible even in the vacuum.
It will be interesting to use this formalism to study 
the rotating detector.

\section{Conclusions}

In this paper we discuss the relativistic problem of uniform 
rotation and how this question is related with the puzzle 
of the rotating detector. After this we discuss the response function 
of a particle detector traveling in different world lines interacting 
with a scalar field prepared in two different vacua: the Minkowski 
and the rotating vacuum.
For electrons in storage rings, a residual depolarization has been 
found experimentally. Bell and Leinaas investigate this effect 
using the idea of circular Unruh-Davies effect. We 
propose a alternative 
solution to the rotating detector puzzle and how this will be related 
with depolarization effects in circular accelerators. 

Let use the result that the probability of transition per unit 
proper time depends not only of the world line 
of the ``atom" but also the particular vacuum in which we prepare the 
field to study four different situations:

i) The response function of an inertial detector interacting with 
the field in the Minkowski vacuum;

ii) The response function of the rotating detector
interacting with the field in the Minkowski vacuum;

iii) The response function of an inertial detector interacting 
with the field in the rotating vacuum;

iv) The response function of the rotating detector
interacting with the field in the rotating vacuum.

The same kind of analysis in a different situation was given by 
Pinto Neto and Svaiter \cite{Nelson}. 
The case (i) gives the usual result that an inertial detector 
in its ground state interacting with the field in the Minkowski 
vacuum does not excite. It is clear that the situation (iv) will 
give the same result.
The case (ii) can be produced in a laboratory. The case (iii) is 
more involved. How to produce the rotating vacuum? A possible 
solution is to use the ideas developed by Denardo 
and Percacci \cite{Denardo} and also Manogue \cite{Manogue}.
This second author consider the case of rotating boundaries to push the 
vacuum around. Note that we are dealing with a Casimir rotating vacuum. 
Is it possible to create some kind of rotating 
vacuum? If the answer is positive we conjecture that the situation 
(iii) will give 
the same response function as situation (ii).

It will be of interest to explore the consequences of this paper, in 
particular to examinate some interesting astrophysical situations. For 
example, the origin of non-thermal radio-frequency in the Universe can be 
explained by the mechanism of synchroton radiation? \cite{Alfin}. Some 
authors discussed the metric of a spinning cosmic string \cite{Mazur}.
We conjecture that electrons and positrons in the neighbourhood of 
such objects must emit synchroton radiation. On the same grounds we 
conjecture that any rotating astrophysical object 
(sppining pulsars for example 
\cite{Taylor}) with a cloud of electrons and positrons 
is a source of synchroton radiation. We can attempt to
justify our conjecture using the well known 
result that the radiation emited by a pulsar has a high 
degree of polarization.
This fact suggest that the mechanism is similar to the
one that generates 
synchroton radiation.

Before finish we would like to made some coments concerning the Sagnac's 
effect. This is the optical analogue of the Foulcault pendulum. In the 
Sagnac's experiment the apparatus device rotates, and the optical 
experiment can determine the rotation of the frame relative to an inertial 
frame. This shows the diference between inertial and the rotating 
(non-inertial) frame. For inertial frames it is impossible to determine the absolute velocity of the apparatus. In the case of the rotating frame the 
angular velocity can be obtained. Our criticism of this scheme is 
the following: to measure the proper spatial line element (in the 
rotating frame) we have to measure the time taken by the light 
signal between an emision and also absorption from atoms. 
The connection with 
the detector puzzle shows how is intricate the analysis. 

In conclusion, in this paper we have attemp to discuss the consequences of 
assuming that a rotating frame is a Lorentz frame of reference.
If we second quantize a scalar field in this frame we show that,
{\it the Minkowski and a rotating vacuum are not the same}. 
Although the Bogoliubov coefficients $\beta_{k\nu}$ 
between the inertial and the non-inertial modes are non-zero it is very 
difficult to calculate them.
We are forced to admit that we fail to finish our interprize since 
we meet a basic difficulty to calculate the spectral density of 
rotating particles in the Minkowski vacuum. Is it possible to go further?

\section{Acknowledgement}

We would like to thank V.Mostepanenko, N.Pinto Neto, I. Dami\~ao Soares,
B.F.Svaiter and F.S.Nogueira for valuable comments. We are also 
grateful to L.H.Ford for several 
valuable conversations and P.C.W.Davies for encouragement.  
This paper was supported by Conselho Nacional de Desenvolvimento 
Cient\'{\i}fico e Tecnol\'ogico (CNPq) do Brazil.

\section*{Appendix A}

In this appendix we will present the solution of Eq.(\ref{51}):
$$
\left[\frac{\partial^{2}}{\partial r^{2}}+
(\frac{1}{r}-4i\omega\Omega\theta r)\frac{\partial}{\partial r}+
\frac{1}{r^{2}}\frac{\partial^{2}}{\partial\theta^{2}}+
(q^{2}-4i\omega\Omega\theta)\right]f(r,\theta)=0.
$$

Let us define $g(r,\theta)$ by the following equation:
$$
f_{\mu}(r,\theta) = e^{i\mu\theta}g_{\mu}(r,\theta).
$$

A direct substitution gives the equation for $g(r,\theta)$:
$$
\left[\left(\frac{\partial^{2}}{\partial r^{2}}+
\frac{1}{r}\frac{\partial}{\partial r} 
-\frac{\mu^{2}}{r^{2}} + q^{2}\right)
+\frac{1}{r^{2}}\left(\frac{\partial^{2}}{\partial\theta^{2}} + 2i\mu\frac{\partial}{\partial\theta}\right)
-\sigma\theta\left( r \frac{\partial}{\partial r}+1 \right)\right]g_{\mu}(r,\theta)=0,
$$
where $\sigma = 4i\omega\Omega$. Define the 
new quantity $y=qr$ and $l = \sigma / q^{2}$ the equation
becomes
$$
\left[\left(\frac{\partial^{2}}{\partial y^{2}}+
\frac{1}{y}\frac{\partial}{\partial y} 
-\frac{\mu^{2}}{y^{2}} + 1\right)
+\frac{1}{y^{2}}\left(\frac{\partial^{2}}{\partial\theta^{2}} + 2i\mu\frac{\partial}{\partial\theta}\right)
-l\theta\left(y \frac{\partial}{\partial y} + 1\right)\right]g_{\mu}(y,\theta)=0.
$$

There appear to be no way of solve the above equation exactly. 
Consequently 
let us try a perturbative solution given by 
$$
g_{\mu}(y,\theta) = J_{\mu}(y) + \sum_{k=1}^{\infty}l^{k}P^{(k)}_{\mu}(y,\theta).
$$
By considering only the first order term in the above expansion  and for simplicity using the 
notation $P^{(1)}_{\mu}(y,\theta)\equiv P_{\mu}(y,\theta)$ we obtain:
$$
\left(\frac{\partial^{2}}{\partial y^{2}}+
\frac{1}{y}\frac{\partial}{\partial y} 
-\frac{\mu^{2}}{y^{2}} + 1\right)P_{\mu}(y,\theta)
+\frac{1}{y^{2}}\left(\frac{\partial^{2}}{\partial\theta^{2}} + 
2i\mu\frac{\partial}
{\partial\theta}\right)P_{\mu}(y,\theta)
-\theta\left(y \frac{\partial}{\partial y} + 1\right)J_{\mu}(y)=0.
$$
Defining 
$$
\frac{1}{2}y^{3}J_{\mu-1}(y) + y^{2}J_{\mu}(y) - 
\frac{1}{2}y^{3}J_{\mu+1}(y) = h(y),
$$
we get:
$$
\left[\left(y^{2}\frac{\partial^{2}}{\partial y^{2}}+
y\frac{\partial}{\partial y} 
-\mu^{2} + y^{2}\right)  + \frac{\partial^{2}}{\partial\theta^{2}} + 2i\mu\frac{\partial}
{\partial\theta}\right]P_{\mu}(y,\theta) = \theta h(y).
$$
It is possible to use the Green\rq s  functions method to 
find the general 
solution for
$P_{\mu}(y,\theta)$. Thus,
$$
P_{\mu}(y,\theta) = P^{(0)}_{\mu}(y,\theta) + \int {\rm d}\theta^{'}\int {\rm d}y^{'} 
G(y,\theta ; y^{'},\theta^{'})\theta^{'}h(y^{'}),
$$
where $P^{(0)}_{\mu}(y,\theta)$ is the solution of the 
homogeneous equation, and 
$G(y,\theta ; y^{'},\theta^{'})$ satisfy 
$$
\left[\left(y^{2}\frac{\partial^{2}}{\partial y^{2}}+
y\frac{\partial}{\partial y} 
-\mu^{2} + y^{2}\right)  + \frac{\partial^{2}}{\partial\theta^{2}} + 2i\mu\frac{\partial}
{\partial\theta}\right]G(y,\theta ; y^{'},\theta^{'}) = \delta(y-y^{'})\delta(\theta - \theta^{'}).
$$
It is straightforward to find the solution of the homogeneous equation 
using separation of 
variables method defining:
$$
P^{(0)}_{\mu}(y,\theta) = e^{i\lambda\theta}Q^{(0)}_{\mu}(y).
$$
Then, 
$$
Q^{(0)}_{\mu}(y) = J_{\mu+\lambda}(y).
$$
Finally the general solution is given by:
\begin{eqnarray}
f_{\mu}(y,\theta) &= & e^{i\mu\theta}\left[J_{\mu}(y)+ 
l e^{i\lambda\theta}J_{\mu+\lambda}(y)\right]
\nonumber\\
&+ &\frac{l}{2}\int {\rm d}\theta^{'}\int {\rm d}y^{'} 
G(y,\theta ; y^{'},\theta^{'})\theta^{'} \left[y^{'3}J_{\mu-1}(y^{'}) + 2y^{'2}J_{\mu}(y^{'}) -
 y^{'3}J_{\mu+1}(y^{'})\right]\nonumber
\end{eqnarray}

\section*{Appendix B}

An orthonormal set is defined through a scalar product in the 
vector space of the solutions of some equation of motion. In the case of 
Klein-Gordon field this scalar product is Hermitian but not positive 
definite. Let be $f(x)$ and $g(x)$ two elements of $F$, where 
$F$ is the vector space of the solutions of the Klein-Gordon 
equation with 
the scalar product defined by
$$
(f,g)=-i\int_{\Sigma}\sqrt{-g}{\rm d}\Sigma^{\mu}\left[f(x)
(\partial_{\mu}g^{*}(x))
-(\partial_{\mu}f(x))g^{*}(x)\right]
$$
where ${\rm d}\Sigma^{\mu}=\eta^{\mu}{\rm d}\Sigma$ with $\eta^{\mu}$ 
a future directed unit vector orthogonal to the space-like 
hypersurface $\Sigma$ and $d\Sigma$ is the volume element 
in $\Sigma$. An orthonormal set $(u_{k},u_{k}^{*})$ is said to be 
complete if every solution $f(x)$ of $F$ can be written as 
$$
f(x)=\sum_{k}a_{k}u_{k}(x)+b_{k}u^{*}_{k}(x)
$$
where the coeficients $a_{k}$ and $b_{k}$ are given by 
$$
a_{k}=(u_{k},f)
$$ 
and
$$
b_{k}=-(u^{*}_{k},f).
$$ 
Let $G$ be a subset of $F$. If$(v_{j},v_{j}^{*})$ and $(u_{i},u_{i}^{*})$
are two orthonormal sets such that the expand every element of $G$, 
then they
are called equivalents. In this case 
$$
v_{j}(x)=\sum_{i}\alpha_{ji}u_{i}(x)+\beta_{ji}u_{i}^{*}(x)
$$ 
and 
$$
u_{i}(x)=\sum_{j}\alpha_{ji}^{*}v_{j}(x)-\beta_{ji}v_{j}^{*}(x).
$$ 
They are said to be complete only if $F=G$.
The quantum field $\varphi(x)$ can be expanded using either 
of the two complete sets $(u_{i},u_{i}^{*})$ or $(v_{j},v_{j}^{*})$
that would lead to two different vacua $|0>$ and $|0'>$ respectively.
When $\sum_{ij}|\beta_{ij}|^{2}$ converges, the representations 
are said to be unitarily equivalent. If it diverges they are non-unitarily 
equivalents and they are not related to any unitary operator in the Fock 
space \cite{Miransky,Umezawa}.

\end{document}